\documentclass[11pt]{amsart}
\usepackage{latexsym,amssymb,amsmath,amscd,amsthm}
\numberwithin{equation}{section}
\theoremstyle{remark}
\newtheorem{theorem}{{\bf THEOREM}}[section]

\newtheorem{corollary}{{\bf COROLLARY}}[section]

\newtheorem{proposition}{{\bf PROPOSITION}}[section]
\newcommand{\bq}{\begin{equation}}
\newcommand{\bea}{\begin{array}}
\newcommand{\eea}{\end{array}}

\newcommand{\ga}{\alpha}
\newcommand{\gep}{\epsilon}
\newcommand{\gD}{\Delta}

\newcommand{\gb}{\beta}

\newcommand{\mf}{\mathfrak}
\newcommand{\mc}{\mathcal}

\newcommand{\go}{\omega}
\newcommand{\gO}{\Omega}

\newcommand{\gd}{\delta}
\newcommand{\pp}{\partial}

\newcommand{\tl}{\tilde}

\newcommand{\bs}{\blacksquare}

\newcommand{{\DDD}}{D\!\!\!\!\!\!-}

\title{UNCERTAINTY, TRAJECTORIES, and DUALITY}
\author{Robert Carroll\\University of Illinois, Urbana, IL 61801}

\date{August, 2003\thanks{email: rcarroll@math.uiuc.edu}}

\dedicatory
{To a wood nymph in an enchanted forest}

\begin{document}

\bibliographystyle{plain}

\begin{abstract}
First we show explicitly how uncertainty can arise in a trajectory representation.
Then we show that the formal utilization of the WKB
like hierarchy structure of dKdV in the description of $(X,\psi)$ duality
does not encounter norm constraints.
\end{abstract}

\maketitle





\section{BACKGROUND}
\renewcommand{\theequation}{1.\arabic{equation}}
\setcounter{equation}{0}

In a previous paper \cite{c3} (working with stationary states and $\psi$
satisfying the Schr\"odinger equation (SE) ${\bf
(A0)}\,\,-(\hbar^2/2m)\psi''+V\psi=E\psi$) we suggested that the notion of
uncertainty in quantum mechanics (QM) can be phrased as incomplete information. 
The background theory here is taken to be the trajectory theory of
Bertoldi-Faraggi-Matone-Floyd (cf.
\cite{b1,c1,c2,ch,f1,f2,f3,f4}).  The idea in \cite{c3} was simply that Floydian
microstates satisfy a third order quantum stationary Hamilton-Jacobi equation
(QSHJE)
\bq\label{1.1}
\frac{1}{2m}(S_0')^2+{\mf W}(q)+Q(q)=0;\,\,Q(q)=\frac{\hbar^2}{4m}\{S_0;q\};
\end{equation}
$${\mf W}(q)=-\frac{\hbar^2}{4m}\{exp(2iS_0/\hbar);q\}\sim V(q)-E$$
where ${\bf (A1)}\,\,\{f;q\}=(f'''/f')-(3/2)(f''/f')^2$ is the Schwarzian and
$S_0$ is the Hamilton principle function.  If one recalls that the EP of 
Faraggi-Matone can only be implemented when $S_0\ne const$ one may think of
${\bf (A2)}\,\,\psi=Rexp(iS_0/\hbar)$ with $Q=-\hbar^2 R''/2mR$ and $(R^2S_0')'=0$
where ${\bf (A3)}\,\,S_0'=p$ and $m_Q\dot{q}=p$ with $m_Q=m(1-\pp_EQ)$ and 
$t\sim\pp_ES_0$.  Thus microstates require three initial or boundary conditions
in general to determine $S_0$ whereas the SE involves only two such conditions.
Hence in dealing with the SE in the standard QM Hilbert space formulation one is
not using complete information about the ``particles" described by microstate
trajectories.  The price of underdetermination is then uncertainty in $q,p,t$ for
example.  In the present note we will make this more precise and add further
discussion.

\section{SOME CALCULATIONS}
\renewcommand{\theequation}{2.\arabic{equation}}
\setcounter{equation}{0}

It is shown in \cite{f2} that one has generally a formula
\bq\label{3.18}
e^{2iS_0(\gd)/\hbar}=e^{i\ga}\frac{w+i\bar{\ell}}{w-i\ell}
\end{equation} 
($\gd\sim (\ga,\ell)$) with three integration constants, $\ga,\ell_1,\ell_2$ where
$\ell=\ell_1+i\ell_2$ and $w\sim \psi^D/\psi\in{\bf R}$.
Note $\psi$ and $\psi^D$ are linearly independent solutions of the SE and one
can arrange that $\psi^D/\psi\in{\bf R}$ in describing any situation.
Here $p$ is determined by the two constants in $\ell$ and has a form
\bq\label{3.19}
p=\frac{\pm\hbar\gO\ell_1}{|\psi^D-i\ell\psi|^2}
\end{equation}
(where $w\sim \psi^D/\psi$ above and $\gO=\psi'\psi^D-\psi(\psi^D)'$).
Now let p be determined exactly with $p=p(q,E)$ via the Schr\"odinger equation and
$S_0'$.  Then $\dot{q}=(\pp_Ep)^{-1}$ is also exact so $\gD
q=(\pp_Ep)^{-1}(\tau)\gD t$ for some $\tau$ with $0\leq \tau\leq t$ is exact (up to
knowledge of $\tau$).  Thus given the wave function
satisfying {\bf (A0)}, 
with two boundary conditions at $q=0$ say to fix
uniqueness, one can create a probability density $|\psi|^2(q,E)$ and
the function
$S_0'$.  This determines $p$ uniquely
and hence $\dot{q}$.  The additional constant needed for $S_0$ 
appears in \eqref{3.18} and we can write $S_0=S_0(\ga,q,E)$ since
from \eqref{3.18} one has ${\bf (A4)}\,\,S_0-(\hbar/2)\ga=-(i\hbar/2)log(\gb)$
and 
$\gb=(w+i\bar{\ell})/(w-i\ell)$ with $w=\psi^D/\psi$ is to be considered as known
via a determination of suitable $\psi,\,\psi^D$.  Hence $\pp_{\ga}S_0=
-\hbar/2$ and consequently ${\bf (A5)}\,\,\gD S_0\sim\pp_{\ga}S_0\gd\ga=-(\hbar/2)
\gD\ga$ measures the indeterminacy in $S_0$.
\\[3mm]\indent
Let us expand upon this as follows.  Note first that the determination of constants 
necessary to fix $S_0$ from the QSHJE is not usually the same as that
involved in fixing $\ell,\,\bar{\ell}$ in \eqref{3.18}.  In paricular differentiating
{\bf (A4)} in q one gets
\bq\label{3.20}
S_0'=-\frac{i\hbar\beta'}{\beta};\,\,\gb'=-\frac{2i\Re\ell w'}{(w-i\ell)^2}
\end{equation}
Since $w'=-\gO/\psi^2$ where $\gO=\psi'\psi^D-\psi(\psi^D)'$ we get ${\bf
(A6)}\,\,
\gb'=-2i\ell_1\gO/(\psi^D-i\ell\psi)^2$ and consequently
\bq\label{3.21}
S_0'=-\frac{\hbar\ell_1\gO}{|\psi^D-i\ell\psi|^2}
\end{equation}
which agrees with p in \eqref{3.19} ($\pm\hbar$ simply indicates direction).  We
see that e.g.
$S_0(x_0)=i\hbar\ell_1\gO/|\psi^D(x_0)-i\ell\psi(x_0)|^2=f(\ell_1,\ell_2,x_0)$
and $S_0''=g(\ell_1,\ell_2,x_0)$ determine the relation between $(p(x_0),
p'(x_0))$ and $(\ell_1,\,\ell_2)$ but they are generally different numbers.  In
any case, taking
$\ga$ to be the arbitrary unknown constant in the determination of $S_0$, we have
$S_0=S_0(q,E,\ga)$ with ${\bf (A7)}\,\,q=q(S_0,E,\ga)$ and $t=t(S_0,E,\ga)=
\pp_ES_0$ (emergence of time from the wave function - ?).  One can then write e.g.
${\bf (A8)}\,\,
\gD q=(\pp q/\pp S_0)(\hat{S}_0,E,\ga)\gD S_0=(1/p)(\hat{q},E)\gD
S_0=-(1/p)(\hat{q},E)(\hbar/2)\gD\ga$ 
(for intermediate values ($\hat{S}_0,\,\hat{q}$)) leading to
\begin{theorem}
With p determined uniquely by two ``initial" conditions so that $\gD p$ is determined
and q given via \eqref{3.18} we have from {\bf (A8)} the inequality ${\bf
(A9)}\,\,
\gD p\gD q=O(\hbar)$ which resembles the Heisenberg uncertainty relation.
\end{theorem}
\begin{corollary}
Similarly ${\bf (A10)}\,\,\gD t=(\pp t/\pp S_0)(\hat{S}_0,E,\ga)\gD S_0$ for some
intermediate value $\hat{S}_0$ and hence as before ${\bf (A11)}\,\,
\gD E\gD t=O(\hbar)$ ($\gD E$ being precise).
\end{corollary}

\section{EMBELLISHMENT FOR ENHANCED DKDV}
\renewcommand{\theequation}{3.\arabic{equation}}
\setcounter{equation}{0}

In \cite{c1,c2,ch} we developed an enhanced dispersionless KdV theory to deal
with the $(X,\psi)$ duality of Faraggi-Matone \cite{f1} for the SE {\bf (A0)}.
The fact that connection of such diverse equations as SE and KdV could arise is
simply based on the fact that they both share a second order differential
equation.  We want to indicate here more precisely this common feature
and will summarize some material from \cite{c1,c2,ch,c5,c6,c7,c9} in the process.

\subsection{$(X,\psi)$ DUALITY}

Consider a Schr\"odinger equation ${\bf
(A12)}\,\,[-(\hbar^2/2m)\pp_X^2+V(X)]\psi=E\psi$ where $E$ is in the spectrum
(V and E are real and large X is used for reasons indicated below).  Then let
$\psi^D$ be a second linearly independent solution of {\bf (A12)} and define a
prepotential
${\mf F}_E={\mf F}$ via ${\bf (A13)}\,\,\psi^D=\pp{\mf F}(\psi)/\pp\psi$ with
${\bf (A14)}\,\,\pp_X{\mf F}=\psi^D\pp_X\psi$ following \cite{f1,f2}. 
One expects ${\bf (A15)}\,\,
{\mf F}=(1/2)\psi\psi^D+G$ with 
\bq\label{1}
\pp_X{\mf
F}=\psi^D\pp_X\psi=\frac{1}{2}(\pp_X\psi)\psi^D+(1/2)\psi\pp_X\psi^D+G_X\Rightarrow
G_X=\frac{1}{2}(\psi^D\pp_X\psi-\psi\pp_X\psi^D)=\frac{1}{2}W
\end{equation}
where the Wronskian $W$ is constant.  Hence take $G=(1/2)WX+c$ with $c=0$
permitted to get ${\bf (A16)}\,\,{\mf F}=(1/2)\psi\psi^D+(1/2)WX$. 
We will consider situations where $\psi^D\sim\bar{\psi}$ with $V$ and $E$ real,
and first, changing a minus sign in \cite{c1,c2,ch} to plus for a neater notation,
we scale W via
${\bf (A17)}\,\,W=\psi'\bar{\psi}-\psi\bar{\psi}'=2i/\gep$ where $\gep\sim
\hbar/\sqrt{2m}$. (cf. Remark 3.4 a description of scaling and
normalization).  Then,  defining
${\bf (A18)}\,\,\phi=\pp{\mf F}/\pp\psi^2=\bar{\psi}/2\psi$ one has a Legendre
transform pair
\bq\label{3}
-\frac{X}{i\gep}=\psi^2\pp_{\psi^2}{\mf F}-{\mf F};\,\,-{\mf F}=\phi\frac
{X_{\phi}}{i\gep}-\frac{X}{i\gep}
\end{equation}
where $(X_{\phi}/i\gep)=\psi^2$ must be stipulated.  
Thus ${\bf (A20)}\,\,{\mf F}=(1/2)\psi\bar{\psi}-(X/i\gep)$, where it is
interesting to note that $X/\gep=x$ in Section 3.1.  This leads to
${\bf (A21)}\,\,
{\mf F}_{\psi\psi\psi}=\frac{(E-V)}{4}
({\mf F}_{\psi}-\psi{\mf F}_{\psi\psi})^3$,
which agrees with previous calculations in \cite{c1,c2,ch} when $\gep\to-\gep$.
\\[3mm]\indent
{\bf REMARK 3.1.}
Equation {\bf (A21)} can be written out in a different manner in terms of ${\mf
F}'=\pp{\mf F}/\pp X$ in the form
${\bf (A22)}\,\,
\gep^2{\mf F}'''-2V'\left({\mf F}+\frac{X}{i\gep}\right)+4(E-V)\left({\mf F}
+\frac{1}{i\gep}\right)=0$.
This resembles a classical Gelfand-Dickey resolvant equation from soliton theory
(cf. \cite{d1}).  Indeed the resolvant equation arises in the form
${\bf (A23)}\,\,R'''+4uR'+2u'R+4z^2R'=0$ in reference to a Lax operator
$L=\pp^2+u$ and R is the formal analogue of the diagonal of the kernel of a
resolvant (i.e. an operator inverse to $\pp^2+u+z^2$ corresponding to a Green's
function.  Such an operator can be constructed as the product
of two solutions of $(\pp^2+u+z^2)\psi=0$ (i.e. $R=\psi_1\psi_2$ which in the
present situation refers to $R=\psi\bar{\psi}$).  Thus $\psi\bar{\psi}=\Xi\sim
2{\mf F}+(2X/i\gep) =2({\mf F}+(X/i\gep))$ with ${\bf (A12)}\sim {\bf (A12')}\,\,
\psi''-(1/\gep^2)V\psi+(1/\gep^2)E\psi=0$ so $u\sim -(V/\gep^2)$ and $z^2\sim 
E/\gep^2$ with ${\mf F}'''=\Xi'''$.  Hence {\bf (A22)} amounts to $\Xi\sim R$
with 
\bq\label{10}
\Xi'''-\frac{2V'}{\gep^2}\Xi+\frac{4(E-V)}{\gep^2}\Xi'=0\equiv
\Xi'''+2u'\Xi+4z^2\Xi'+4u\Xi'=0
\end{equation}
This gives us a first direct connection between $(X,\psi)$ duality and the
KdV theory based on the fact that a very special second order differential
equation ${\bf (A12)} \sim {\bf (A12')}$
appears in both situations ($\gep^2\sim \hbar^2/2m$ - see below).  Evidently such
connections based on properties of an operator $\pp^2+u$ are perfectly natural
and mathematically well founded;
in particular we will use techniques from the theory of ${\bf (A12')}$ to describe
quantities arising via {\bf (A12)} in $(X,\psi)$ duality.  The fact that some of the
techniques arise from KdV (or dKdV) theory via tau functions, Lax operators, etc. is
purely incidental (and fortuitous).$\hfill\bs$

\subsection{DISPERSIONLESS THEORY FOR KP and KDV}

We assume the standard KP and KdV theory is known (cf. \cite{ch,c5,d1}).
Then one can think of fast and slow variables with $\epsilon x=X$ and $\epsilon
t_n= T_n$ so that $\partial_n\to\epsilon\partial/\partial T_n$ and $u(x,t_n)
\to U(X,T_n)$ to obtain from the KP equation $(1/4)u_{xxx}+3uu_x
+(3/4)\partial^{-1}\partial^2_2u=0$ the equation $\partial_TU=3
U\partial_XU+(3/4)\partial^{-1}(\partial^2U/
\partial T_2^2)$ when $\epsilon\to 0$ ($\partial^{-1}\to(1/\epsilon)
\partial^{-1}$).  In terms of hierarchies the theory can be built around the
Lax operator $L=\partial+\sum_1^{\infty}u_{n+1}\partial^{-n}$.
Then writing $(t_n)$ for
$(x,t_n)$ (i.e. $x\sim t_1$ here) consider
${\bf (A23)}\,\,
L_{\epsilon}=\epsilon\partial+\sum_1^{\infty} u_{n+1}(\epsilon,T)
(\epsilon\partial)^{-n}$.
Now one assumes $u_n(\epsilon,T)=U_n(T)+O(\epsilon)$, etc. and 
sets (recall $L\psi=\lambda\psi$)
\bq\label{YB}
\psi=\left[1+O\left(\frac{1}{\lambda}\right)\right]exp
\left(\sum_1^{\infty}\frac{T_n}
{\epsilon}\lambda^n\right)=exp\left(\frac{1}
{\epsilon}S(T,\lambda)+O(1)\right);
\end{equation}
where ${\bf (A24)}\,\,
\tau=exp\left(1/\epsilon^2)F(T)+O\left(1/\epsilon)\right)
\right)$.
Note for the approximation of potentials one
assumes e.g. $v=v(x,t_i)\to v(X/\epsilon,T_i/\epsilon)
=V(X,T_i)
+O(\epsilon)$.  This is
standard in dispersionless KP = dKP and certainly realized
by quotients of homogeneous polynomials for example. 
In fact it is hardly a restriction since we are primarily interested here in
the X dependence and given e.g. $F(X)=\sum_0^{\infty}
a_nX^n$ consider $\tilde{F}(x,t_i)=a_0+\sum_1^{\infty}(x^n/\prod_2^{n+1}
t_i)$.  Then $\tilde{F}(X/\epsilon,T_i/\epsilon)=a_0+\sum_1^{\infty}
(X^n/\prod_2^{n+1}T_i)$ and one can choose the $T_i$ recursively so that
$1/T_1=a_1,\,\,1/T_1T_2=a_2,\cdots$, leading to $F(X)=\tilde{F}(X,T_i)$. 
Thus one can work with a $t\sim T$ dependent theory and eventually capture the
original $F(X)$ for a stationary theory by specializing and freezing the $T_i$. 
\\[3mm]\indent
{\bf REMARK 3.2.}
We recall also that $\partial_nL=[B_n,L],\,\,B_n=L^n_{+},
\,\,L\psi=\lambda\psi$, and in terms of tau functions
$\psi=\tau(T-(1/n\lambda^n))exp[\sum_1^{\infty}T_n\lambda^n]/
\tau(T)$.  Putting in the $\epsilon$ and using $\partial_n$ for
$\partial/\partial T_n$ now, with $P=S_X$, one obtains
${\bf (A25)}\,\,
\lambda=P+\sum_1^{\infty}U_{n+1}P^{-n}$ and 
$P=\lambda-\sum_1^{\infty}W_i\lambda^{-1}$ (when $\gep\to 0$)
There are then many beautiful formulas for dKP (see e.g. \cite{ch}) but we emphasize
here that in constrast to dKdV we are not going to let $\gep\to 0$; instead we use
it to balance terms below.$\hfill\bs$
\\[3mm]\indent
Now look at the dispersionless theory for KdV based on $k$ where $\lambda^2
\sim(\pm ik)^2=-k^2$. When $\gep\to 0$ one obtains for $P=S_X,\,\,P^2+q=-k^2$, and
we write
${\mc P}=(1/2)P^2+p=(1/2)(ik)^2$ with $q\sim 2p\sim 2u_2$.  One has
$\partial k/\partial T_{2n}=\{(ik)^{2n},k\}=0$ and from $ik=P(1+qP^{-2})^
{1/2}$ we obtain
\bq
ik=P\left(1+\sum_1^{\infty}{\frac{1}{2}\choose m}q^mP^{-2m}\right)
\label{YR}
\end{equation}
($\gep\to 0$ here and we use $+ik$ since it is appropriate later).
We refer now to \cite{ch,c6,c7,t1} for 
dispersionless KP ($=$ dKP) and consider
here $\psi=exp[(1/\epsilon)S(X,T,\lambda)]$ (note ${\bf
(A26)}\,\,L^2=\pp_x^2-v\to \gep^2\pp_X^2-V$).  Thus $P=S'=S_X$ with $P^2=V-E$ 
(when $\gep\to 0$)
and $E=\pm\lambda^2$ real will involve us in a KdV situation and
some routine calculation yields;
recall $X_{\psi}=1/\psi'$ with $\psi'=(P/\epsilon)\psi$ (cf. \cite{ch,c7}).  This
leads to
\bq
\Im{\mf F}=\frac{X}{\epsilon};\,\,\Re{\mf F}=\frac{1}{2}|\psi|^2=
\frac{1}{2}e^{\frac{2}{\epsilon}\Re S}=-\frac{1}{2\Im P}
\label{AGQ}
\end{equation}
In the present situation $|\psi|^2=exp[(2/\epsilon)\Re S]$ and $2\phi=
exp[-(2i/\epsilon)\Im S]$ can play the roles of independent variables
and we see that ${\bf (A27)}\,\,
|\psi|^2\Im P=1$ while  
$\psi^2\phi=(1/2)|\psi|^2$.
Now note that for $L=\partial+\sum_1^{\infty}u_i
\partial^{-i},\,\,L^2_{+}=\partial^2+2u_1$, and $u_1=\partial^2log(\tau)$
where $\tau$ is the famous tau function.  From {\bf (A26)}
this implies $v=-2\partial^2log(\tau)$ here, from which $V=-2F_{XX}$ for
$\tau=exp[(1/\epsilon^2)F+O(1/\epsilon)]$ 
in the dispersionless theory (cf. \cite{ch,c7}). 
Then writing out the Gelfand-Dickey resolvant equation {\bf (A22)}
yields
\bq
\epsilon^2{\mf F}'''+\left({\mf F}'+\frac{1}{i\epsilon}\right)(8F''
+4E)+4F'''\left({\mf F}+\frac{X}{i\epsilon}\right)=0
\label{AKQ}
\end{equation}
which provides a relation between $F$ and ${\mf F}$.
This is interesting since F plays the role of a prepotential or free
energy in the dKdV theory (cf. \cite{c1,c2,ch}).  We recall below how to embellish
all this with an $\gep$-modification (or enhancement) of dKP and dKdV.  Note that
there is no neglect of
$O(\gep)$ terms in
\eqref{AKQ} and $X/i\gep=x$ should be well defined which suggests a scale dependence
in QM and perhaps the emergence of space from the wave function.
\\[3mm]\indent
Next one has $\Im {\mf F}=X/\epsilon$ and 
$|\psi|^2=exp[(2/\epsilon)\Re S]$.  In order to have $|\psi|^2\leq 1$
as a fundamental variable in the $(X,\psi)$ theory some control over
$\Re S$ is needed and this problem is resolved below in Remark 3.5 and Proposition
3.1.  One notes that dKdV  (with $\gep\to 0$ here) involves $\lambda=\pm
ik=P(1+
\sum_1^{\infty}U_mP^{-2m})$ with $\lambda^2=-k^2$ real for $k$ real  
(as in (3.6)).  In order to satisfy $|\psi|^2\Im P=1$ we will
want $\lambda=ik$ so that $Q=\Im P$ is positive. 
The $U_m$ are real so $P=iQ$
corresponds to $k$ real and 
$(ik)^{2n+1}_{+}$ will be purely imaginary (only ``times"
$t_{2n+1}$ arise in KdV).  In addition
$\Re S=\Re[\sum_0^{\infty}T_{2n+1}(ik)^{2n+1}+\sum_1^{\infty}S_{j+1}
(ik)^{-j}]=0$ for $k$ real since $S_{j+1}=-\partial_jF/j$ with 
$\partial_{2n}F=0$.  Now this would imply $|\psi|^2=1$, which is absurd,
so we introduce an expansion $S\to\tl{S}=\sum\epsilon^jS^j$ (in particular this takes
into  account the $O(1)$ terms in $\psi=exp[(1/\epsilon)S+O(1)]$).  Note that such
$O(1)$ terms (and others) arise quite naturally from the vertex operator equation
(VOE) via 
$log\psi=(S/\epsilon)+O(1)= log\tau[\epsilon,T_n-(\epsilon/
n\lambda^n)]-log\tau+\sum_1^{\infty}T_n\lambda^n/\epsilon$ with $log\tau
=(F/\epsilon^2)+O(1/\epsilon)$ and $S_{n+1}^0=-(\partial_nF/n)$.
Thus e.g.
\bq
F\left(T_n-\frac{\epsilon}{n\lambda^n}\right)-F(T_n)=-\epsilon\sum_1^{\infty}
\left(\frac{\partial_nF}{n\lambda^n}\right)+\frac{\epsilon^2}{2}\sum
\left(\frac{F_{mn}}{mn}\right)\lambda^{-m-n}+O(\epsilon^3)
\label{five}
\end{equation}
This leads to first terms of the form $\tl{S}=S^0+\epsilon S^1$ with
\bq
S^1=\frac{1}{2}\sum\left(\frac{F_{mn}}{nm}\right)\lambda^{-m-n};\,\,
S_X\sim P+\frac{\epsilon}{2}\sum\left(\frac{F_{1mn}}{nm}\right)\lambda^
{-m-n}
\label{six}
\end{equation}
Eventually we will want also an embedding $F\to\tl{F}=\sum\epsilon^jF^j$ as well. 
With this  development one finds as a first approximation in $\gep$
\bq
|\psi|^2=e^{2\Re S^1}=exp\left[\Re\sum\left(\frac{F^0_{mn}}{mn}\right)
\lambda^{-m-n}\right]
\label{OF}
\end{equation}
where only terms $F^0_{2m+1,2n+1}\lambda^{-2(m+n)-2}$ arise and these
will be real for $k$ real (so $|\psi|^2\leq 1$ becomes tenable.  We refer to 
Remark 3.4 and Proposition 3.1 where it is shown that this can always be achieved
without constraints.
Thus ${\bf (A28)}\,\,S^0$ and $P^0=S^0_X=iQ$ are
imaginary while $S^1$ and $P^1=S^1_X$ are real.  
\\[3mm]\indent
{\bf REMARK 3.3.}
We note from \cite{f2} that a standard WKB approach to
$(\hbar^2/2m)\psi''+(E-V)\psi=0$ involves
$\psi=Aexp(is/\hbar)$ where $s$ and $A$ are even functions of $\gep$.
This leads to 
${\bf (A29)}\,\,
(s')^2-2m(E-V)=\hbar^2\frac{A''}{A};\,\,2A's'+As''=0$.
Thus ${\bf (A30)}\,\,A=c(s')^{-1/2}$ and the first equation becomes 
${\bf (A31)}\,\,
(s')^2=2m(E-V)-\frac{\hbar^2}{2}\{s;x\};\,\,\{s;x\}=\frac{s'''}{s'}-
\left(\frac{s''}{s'}\right)^2$.
We recall that the $(X,\psi)$ duality theme 
was based on $-\gep^2\psi''+V(X)\psi=E\psi$ with
${\mf F}=(1/2)\psi\bar{\psi}-(X/i\epsilon)$ and 
$\psi=exp[\tilde{S}/\epsilon]$ was employed ($\epsilon=
\hbar/\sqrt{2m}$) with ${\bf (A32)}\,\,\tilde{S}=\sum_0^{\infty}\epsilon^jS^j$
($S^{2j+1}$ real and $S^{2j}$ imaginary for $k$ real, $\lambda=ik$).
This is related (in an expanded dKdV
theory) to $\tl{s}\sim\sqrt{2m}\Im\tilde{S}$ and $log\,A=
(1/\epsilon)\Re\tilde{S}$.  Also since $\psi$ and
$\bar{\psi}$ satisfy the Schr\"odinger equation we know that the square
eigenfunctions $\psi\bar{\psi},\,\,\psi^2$, and $(\bar{\psi})^2$ satisfy
the GD equation.
Further from \cite{f2} a
general solution of {\bf (A31)} is given by
$s'=\pm\sqrt{2m}(a\psi^2+b\bar{\psi}^2 +c\psi\bar{\psi})^{-1}$ where
$\psi,\,\,\bar{\psi}$ are normalized solutions of ${\bf (A32)}$ so we
introduce a constraint 
${\bf (A33)}\,\,\int|\psi|^2dX=1$.
Here one recalls that $|\psi|^2=1/\Im\tilde{P}=exp[(1/\epsilon)\Re\tilde{S}]
=exp[\sum_0^{\infty}\epsilon^{2j}S^{2j+1}]$ and evidently one will have to scale
$\psi\to c\psi$ in order to insure
${\bf (A33)}$; this amounts to a normalization of a BA
function and we will show that this introduces no problems (cf. Remark 3.5).  Now the
GD equation (\ref{10}) for ${\mf F}$ can be written in terms of $\Xi=\psi\bar{\psi}$,
or $\psi^2$, or $\bar{\psi}^2$ as ${\bf (A34)}\,\,
\epsilon^2\Xi'''-4V\Xi'-2V'\Xi+4E\Xi'=0$
so given $s'$ as above one sees the connection via
$s'=c/|\psi|^2$ satisfies {\bf (A31)} while
$|\psi|^2=c/s'$ satisfies {\bf (A34)} (for any c).  We have also shown incidentally
that $|\psi|^2s'=c$ is a consequence of the second order equation ${\bf
(A12)}\sim{\bf (A12')}$ and thus arises from the WKB approach (independently of
$(X,\psi)$ duality).
$\hfill\bs$
\\[3mm]\indent
In \cite{c1,c2,ch} we employed a full connection to dKdV by using the tau function
${\bf (A35)}\,\,\tau=exp[(1/\gep^2)F(T)+O(1/\gep)]$ with ${\bf (A36)}\,\,
V=-2F_{XX}=-2F''$.  Then the GD equation takes the form ${\bf (A37)}\,\,
\gep^2\Xi'''+4F'''\Xi+4E\Xi'+8F''\Xi'=0$.  It is in fact F, with its fundamental
coefficients ${\bf (A38)}\,\,F_{mn}=\pp_m\pp_mF=\pp^2F/\pp T_m\pp T_n$ and their
relations to Hirota theory, which make the enhanced dKdV theory interesting and
useful.  We recall here the expansions ${\bf (A39)}\,\,\tl{S}=\sum
\gep^jS_j,\,\,\tl{F}=\sum \gep^j F_j,\,\,\tl{P}=\tl{S}_X=\sum\gep^jP_j$ which from
the SE $\gep^2\psi''+2F''\psi=-E\psi$ and $\gep^2\psi_{XX}=\gep\tl{P}_X\psi+
\tl{P}^2\psi$ yield
\bq\label{3.6}
\left(\sum_0^{\infty}\gep^jP_j\right)^2+\gep\sum_0^{\infty}\gep^jP_j'+
2\sum_0^{\infty}\gep^jF_j''=-E=-k^2
\end{equation}
(we use sub or superscripts according to notational convenience
and will write $S^0$ in order to distinguish this from our original $S_0\sim s$). 
This leads to
${\bf (A40)}\,\,
P_0^2+2F_0''=-E,\,\,2P_0P_1+P_0'+2F_1''=0,\,\,P_1^2+2P_0P_2+P_1'+2F_2''=0,\cdots$
(there is also another (equivalent) way to determine coefficients indicated in
\cite{c1,c2,ch}).
Take now $F_{2j+1}=0$ (for consistency of equations - cf. \cite{c1,c2,ch}); this
only stipulates what kind of $\gep$ extensions of F are compatible with $(X,\psi)$
duality in an extended sense, namely ${\bf
(A41)}\,\,exp[(2/\gep)\Re\tl{S}]\Im\tl{P}= 1$ (which also appears as an essential
ingredient for using WKB type methods).  Then from
\eqref{3.6} one obtains
${\bf (A42)}\,\,
P_k^2+2\sum_0^{k-1}P_iP_{2k-i}+P'_{2k-1}+2F_{2k}''=0;\,\,
2\sum_0^kP_iP_{2k+1-i}+P'_{2k}=0$
(cf. \cite{c1,c2,ch}).  In particular
${\bf (A43)}\,\,
P_0^2+2F''_0=-E;\,\,2P_0P_1+P'_0=0;\,\,P_1^2+2P_0P_2+P'_1+2F''_2=0$ with
$P'_2+2(P_0P_3+P_1P_2)=0;\cdots$.
Consequently ${\bf (A44)}\,\,P_1=-P'_0/2P_0$ and one can recover all $P_{2j}$ from
E and $F''_{2m}$.  Hence
$\tl{s}'$ is determined by E and the $F''_{2m}$ and this will be a solution of an
enhanced QSHJE.  In terms of construction we note that ${\bf (A45)}\,\,
P_0^2=-2F_0''-E\sim V-E$ and via {\bf (A44)} one has $2P'_0P_1+2P_0P'_1+P''_0=0$
so
${\bf (A46)}\,\,
2P_0P_2=-2F''_2-P'_1-P_1^2=-2F''_2+\frac{1}{2}\left[\left(\frac{P''_0}{P_0}\right)
-\frac{3}{2}\left(\frac{P'_0}{P_0}\right)^2\right]$.
(note the
Schwarzian representation $(1/2)\{S_0;q\}$
in {\bf (A46)}.  In any
event 
$V,\,E,$
and $F''_2$ determine $P_2$.  Continuing one can write
${\bf (A47)}\,\,P_{2k}=f(V,E,F''_{2m}\,\,(1\leq m\leq k))$ and there are no
integration constants involved.  We note that even if $F_2''=0$ there is a
correction term ${\bf (A48)}\,\,P_2=\{S_0;q\}/4P_0$.
\\[3mm]\indent
{\bf REMARK 3.4.}
Let us examine again ${\bf (A49)}\,\,
|\psi|^2\leq 1$ and ${\bf (A57)}\,\,|\psi|^2\Im P=1$.  First {\bf (A50)} is
fundamental to
$(X,\psi)$ duality and in fact to the whole idea of a WKB type expression
$\psi=exp(S/\gep)$.  Thus write the solution of ${\bf
(A51)}\,\,(\gep^2\pp_X^2-V)\psi=-k^2\psi$ as $\psi = Rexp(is/\gep)$ so that
${\bf (A52)}\,\,
\psi'\bar{\psi}=R'R+iR^2s'/\gep$.  Also {\bf (A51)} implies ${\bf (A53)}\,\,
(s')^2+V-E-(\gep^2R''/R)=0$ with $\pp(R^2s')=0$.  In particular ${\bf (A54)}\,\,
R^2s'=c$ (constant) and since $s'\sim\Im S_X=\Im P$ (or $\Im\tl{P}$) we have $|\psi|^2
\Im\tl{P}=c$ by virtue of the WKB formulation.  Note also
from
$(\gep^2\pp_X^2-V)\bar{\psi}=-k^2\bar{\psi}$ one has for $\bar{\psi}=Rexp(-is'/\gep)$ 
the formula $\bar{\psi}'\psi=R'R-(is'R^2/\gep)$.  Combining this with {\bf (A52)}
leads to ${\bf
(A55)}\,\,W=\psi'\bar{\psi}-\bar{\psi}'\psi=2iR^2s'/\gep=(2i/\gep)c$.
Thus the normalization of W involves an integration constant c so one
could say that c is determined by the normalization of W.  Note however that using
$R^2s'=c$ in rewriting the ``quantum potential" $Q=-\gep^2R''/R$ in terms of $s'$ we
have
$2RR's'+R^2s''=0$ with ${\bf (A56)}\,\,
\frac{R''}{R}=-\frac{1}{2}\left[\frac{s'''}{s'}+\frac{3}{2}\left(\frac{s''}{s'}
\right)^2\right]$.
Hence ${\bf (A57)}\,\,Q=(\gep^2/2)\{s;X\}$ and the constant disappears in
constructing the QSHJE - which then requires three ``initial" conditions for
integration.  The important point here is that ${\bf (A58)}\,\,|\psi|^2=R^2=c/s'$
with c a priori an integration constant and in addition c scales the Wronskian as in
{\bf (A55)}.  We could also simply scale $\psi$ via $\psi\to\sqrt{\go}\psi$
to achieve the same result for $\go\sim c$.  This is
interesting since c disappears in the QSHJE and loses its status of integration
constant.  Therefore we can think of c as a scaling factor ($c\sim\go$) in the
projective ray representation of $\psi$.  This does not however fix the theory to
take place on a  sphere of radius $|\psi|^2=\go$ but appears to be a harmless scaling
which can  in fact be used to force
$|\psi|^2\leq 1$ and we will have {\bf (A54)} to measure this.
In any case ${\bf (A59)}\,\,|\psi|^2=exp[2\Re\tl{S}/\gep]=exp[2\sum\gep^{2j}P_{2j+1}]$
and we are free to scale $\psi$ so for
$\psi\to\sqrt{\go}\psi$ with $|\psi|^2\to
\go|\psi|^2$ the requirement is
${\bf (A60)}\,\,
log(\go|\psi|^2)=log|\psi|^2+log\go=log\go+2\sum_0^{\infty}\gep^{2j}P_{2j+1}\leq 0$.
But for $\go<1$ one has $log\go<0$ so given any magnitude for the series
$2\sum_0^{\infty}\gep^{2j}P_{2j+1}$ we can find $\go$ such that $(\go|\psi|^2)\leq
1$.  
$\hfill\bs$
\begin{proposition}
The apparent integration constant c in $R^2s'=c$ can be thought of in terms of an
arbitrary scaling constant $c=\go$ for $\psi$ and hence there need be no concern about
satisfying $|\psi|^2\leq 1$ in the enhanced dKdV theory.
\end{proposition}


\end{document}